\documentclass[
 aps,
prl,
 reprint,
 superscriptaddress,
 nofootinbib,
 amsmath,
 amssymb
]{revtex4-2}

\usepackage{graphicx}
\usepackage{dcolumn}
\usepackage{bm}

\begin{document}

\preprint{APS/123-QED}

\onecolumngrid
\begin{center}

{\large\bfseries
Efficient Proton Relay Orchestrated by Covalent Bond Switching of Active Amino\\
Acids in Protein Channels
\par}

\vspace{1.0em}

{\normalsize
Xiangyu Su,$^{1,*}$
Yuwei Cao,$^{2,*,\dagger}$
Shuyang Zhao,$^{3}$
and Wanlin Guo$^{1,\ddagger}$
\par}

\vspace{0.6em}

{\small\itshape
$^{1}$National Key Laboratory of Mechanics and Control for Aerospace Structures and Key\\
Laboratory for Intelligent Nano Materials and Devices of the Ministry of Education,\\
Institute for Frontier Science, Nanjing University of Aeronautics and Astronautics, Nanjing 210016, China
\par}

{\small\itshape
$^{2}$Innovative Centre for Flexible Devices (iFLEX),\\
Max Planck--NTU Joint Lab for Artificial Senses, School of Materials Science and Engineering,\\
Nanyang Technological University, 50 Nanyang Avenue, 639798, Singapore
\par}

{\small\itshape
$^{3}$School of Aerospace Engineering, Xi'an Jiaotong University, Xi'an 710049, China
\par}

\vspace{0.4em}

{\small
(Dated: June 14, 2026)
\par}

\end{center}
\vspace{-1.2em}

\begin{abstract}
Through systematic mutational simulations of the key site in a proton channel, we find that 13 of the 20 canonical amino acid residues are active for proton transfer through covalent bond switching, whereas the remaining 7 residues, whose side chains terminate in \(sp^3\) hybridized carbon--hydrogen covalent bonds, do not undergo such bond switching and are therefore inactive. All active residues have a negative electrostatic potential extremum at the proton accepting atom and lower energy barriers for proton relay orchestrated by bond switching, whereas the inactive residues have positive electrostatic potential extremum and significantly higher barriers for bond switching. We further find that the active residues tend to be distributed within the pore to mediate proton transfer, while the inactive residues are enriched in the periphery to stabilize the structure. This bond switching activity can also be observed in respiratory complex I. These findings establish a new classification criterion for amino acids based on their covalent bond switching activity, providing insights into how life utilizes the 20 types of amino acids. 

\end{abstract}

\maketitle

\begingroup
\renewcommand{\thefootnote}{\fnsymbol{footnote}}
\footnotetext[1]{These authors contributed equally to this work.}
\footnotetext[2]{Contact author: yuwei.cao@ntu.edu.sg}
\footnotetext[3]{Contact author: wlguo@nuaa.edu.cn}
\endgroup
\setcounter{footnote}{0}

DNA is composed of only four nucleobases, yet encodes the twenty canonical amino acids through the triplet genetic code for protein synthesis~\cite{Crick1961}. These amino acid residues possess diverse side chain structures and physicochemical properties, including charge~\cite{Machin2023}, polarity~\cite{Kamtekar1993,Micheletti1998}, and hydrophobicity~\cite{Rose1985,vanDijk2016,Bianco2017}, but all terminate in hydrogen atoms covalently bonded to the X atom (oxygen/sulfur/nitrogen/carbon) on the residues, forming covalent X(O/S/N/C)–H bonds. The sequential arrangement of amino acids gives rise to proteins with complex structures~\cite{Dudko2011,Dill2012,Yuan2024} and diverse functions, such as mass transport~\cite{Valet2019,Gibby2021,Doyle1998,Kolomeisky2007} and energy conversion~\cite{Toyabe2010,Kampjut2020}, both of which are often coupled to proton transfer, especially in proton channels and bioenergetic enzymes~\cite{Morgan2009,Musset2011,Kim2009,Wu2014}. 

Proteins have stable backbones formed by strong peptide bonds (\(\sim 375\text{--}422~\mathrm{kJ\,mol^{-1}}\))~\cite{Marochkin2012} and disulfide bridges (\(\sim 280~\mathrm{kJ\,mol^{-1}}\) for S--S)~\cite{Roux2010}, and can fold into functional structures through electrostatic interactions (\(\sim 50~\mathrm{kJ\,mol^{-1}}\))~\cite{Zhou2018}, hydrogen bonds (\(\sim 2\text{--}17~\mathrm{kJ\,mol^{-1}}\))~\cite{Perrin1997} and van der Waals forces (\(\sim 0.5\text{--}8~\mathrm{kJ\,mol^{-1}}\))~\cite{Roth1996}. In comparison, the X--H covalent bonds with bond dissociation energies of \(\sim 250\text{--}540~\mathrm{kJ\,mol^{-1}}\)~\cite{Blanksby2003,Treyde2022} are always treated as fixed bonds in structural and functional studies of proteins~\cite{Hollingsworth2018,MacKerell1998,Yukawa2025}. Recently, we showed by quantum dynamics studies that the amino acid residue Glu119 in the human voltage-gated proton channel (hHv1) can mediate proton transfer via O--H covalent bond switching on its carboxyl group (--COOH)~\cite{Cao2024}. It is also noted that the residue His in the M2 proton channel of influenza A virus can mediate proton transfer through bond switching of the imidazole~\cite{Hu2010,Liang2016,Watkins2022}. These examples indicate that some amino acid residues can become active in the presence of protons, which are ubiquitous in cellular environments and living systems. Therefore, a systematic and in-depth understanding of the quantum dynamic interactions between all amino acid residues and the proton becomes a fundamental and urgent issue that needs to be addressed in order to comprehend how life works.

Here, we perform quantum molecular dynamics simulations of systematic mutagenesis of residue Glu119 in the hHv1 proton channel and reveal that the 20 types of amino acid residues can be divided into two groups: 13 active and 7 inactive residues. 
The 13 active residues whose side chains terminate in O/S/N--H and \(sp^2\) C--H \((\mathrm{C}sp^2\)--H) bonds can mediate proton transport through \(\mathrm{X}(\mathrm{O}/\mathrm{S}/\mathrm{N}/\mathrm{C}sp^2\))--H covalent bond switching with low energy barriers \((<20~\mathrm{kJ~mol^{-1}})\), whereas the 7 inactive residues whose side chains terminate in \(sp^3\) C--H \((\mathrm{C}sp^3\)--H) bonds are stable in the cellular environment and mainly support proton transport through the adjacent water molecule wire. 
The activity of amino acid residues is mainly determined by the nucleophilicity and electrostatic potential extremum at the proton accepting atom. 
Interestingly, the active residues preferentially distribute within the pore to mediate proton relay, while the inactive residues mainly distribute in the periphery for structural stabilization. 
We also find this \(\mathrm{X}\)--H bond switching mechanism in respiratory complex I, suggesting a general mechanism for efficient proton transfer in biological systems.

To investigate the proton relay dynamics of different amino acids, we performed quantum mechanics/molecular mechanics (QM/MM) molecular dynamics (MD) simulations [see Supplemental Material (SM)~\cite{SM} Sec.~S1 for details] in hHv1 with Glu119 mutated to each of the remaining 19 types of amino acid residues [Fig.~1(a) and Fig.~S1~\cite{SM}]. 
Glu119 has been shown to mediate proton transfer via the O--H covalent bond switching between the proton and the \(-\mathrm{COOH}\) group on the side chain~\cite{Cao2024}. 
Simulation results indicate that the proton transfer behavior near different amino acid residues depends on the type of their \(\mathrm{X}\)--H bonds in the side chain [Fig.~1(b)]. Residues whose side chains terminate in O/S/N--H and \(\mathrm{C}sp^2\)--H bonds (O/S/N/\(\mathrm{C}sp^2\)--H bonded residues), including Glu, Asp, Ser, Tyr, Thr, Cys, Asn, His, Lys, Trp, Arg, Gln, and Phe, can mediate proton relay through covalent \(\mathrm{X}\)--H bond switching (Fig.~S2~\cite{SM}). 
We therefore define these 13 residues as active residues. 
In contrast, the residues whose side chains terminate in \(\mathrm{C}sp^3\)--H bonds (\(\mathrm{C}sp^3\)--H bonded residues), including Leu, Ala, Val, Gly, Ile, Met, and Pro do not undergo \(\mathrm{X}\)--H bond switching; instead, protons are transferred through adjacent water molecules (Fig.~S3~\cite{SM}). 
These 7 residues are defined as inactive residues.

\begin{figure}[t]
\centering
\includegraphics[width=\columnwidth]{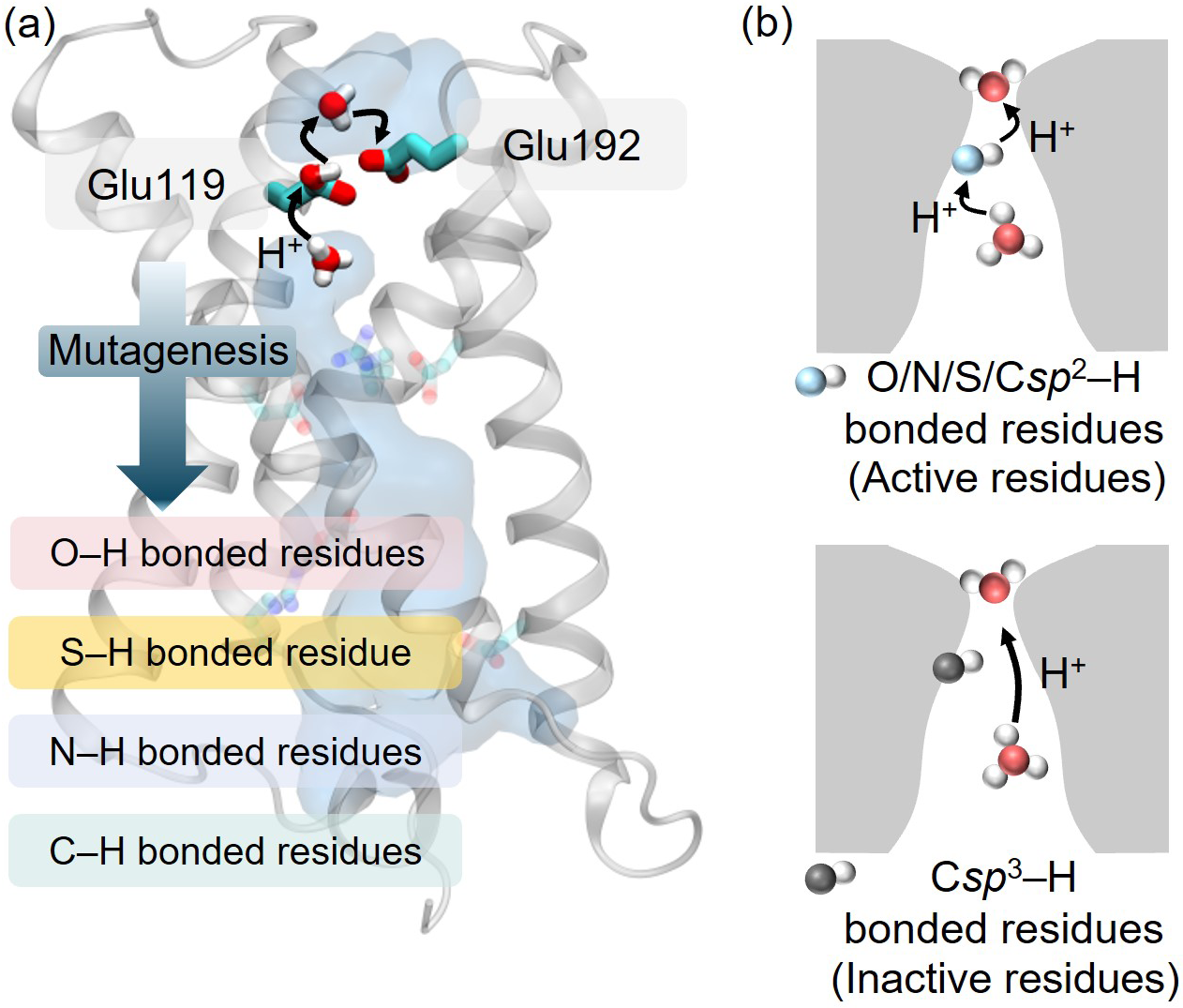}
\caption{
Proton relay via X--H bond switching on active residues in hHv1 proton channel. 
(a) Atomic model of hHv1 protein channel. The blue transparent region denotes the space containing water molecules. Residues involved in proton transfer are shown in licorice, including Glu119 and Glu192. Glu119 was systematically mutated to the remaining 19 canonical amino acids. 
(b) Proton transfer mechanisms of active and inactive residues: active residues (O/S/N/\(\mathrm{C}sp^2\)--H bonded residues) mediate proton relay via X-H bond switching, inactive residues (\(\mathrm{C}sp^3\)--H bonded residues) do not participate in proton transfer, and protons transfer along adjacent water molecules.
}
\label{fig:fig1}
\end{figure}

The \(\mathrm{X}\)--H bond switching processes on active amino acid residues are not always the same even though they contain the same functional group. For example, Glu and Asp containing \(-\mathrm{COOH}\) groups exhibit distinct covalent O--H bond switching pathways. The O--H bond switching in Glu occurs at the same oxygen site but at different oxygen sites in Asp [Figs.~2(a) and 2(b)]. Elements belonging to the same group in the periodic table tend to undergo similar bond switching processes. Notably, the thiol group \((-\mathrm{SH})\) in Cys can also mediate proton relay via S--H bond switching, analogous to the O--H bond switching on residues (Ser, Tyr and Thr) containing hydroxyl groups \((-\mathrm{OH})\) [Fig.~2(c) and Fig.~S2(b)~\cite{SM}]. 
For N--H bonded residues, N--H bond switching generally occurs at a single nitrogen site, as observed in Lys, Trp, Asn, Gln, and Arg (Fig.~S2(d)~\cite{SM}). In contrast, His, which contains an imidazole ring, can mediate proton transfer via concerted N--H bond switching on the two nitrogen sites, where one nitrogen acts as a proton acceptor while the other donates its originally bonded hydrogen atom to a neighboring water molecule. Therefore, His may provide an efficient and long-distance proton relay pathway [Fig.~2(d)]. 

Moreover, proton transport behavior depends not only on the elemental type of \(\mathrm{X}\)--H bonds in residues but also on the hybridization state of \(\mathrm{X}\) atoms, especially for the C--H bonded residues. Most of them are \(\mathrm{C}sp^3\)--H bonded residues, where no bond switching can be observed, and protons are transferred along continuous water wires formed around them (see Fig.~S3~\cite{SM} for details). Specifically, the \(\mathrm{C}sp^3\)--H bonded Met, which terminates in the methyl group \((-\mathrm{CH_3})\), exhibits rotational rearrangement to form a continuous water wire for proton transport [Fig.~2(e)]. In contrast, the \(\mathrm{C}sp^2\)--H bonded Phe mediates proton transfer via \(\mathrm{C}sp^2\)--H bond switching: the incoming proton replaces the original hydrogen on the phenyl \(\mathrm{C}sp^2\)--H site, and the displaced hydrogen is subsequently transferred to a neighboring water molecule [Fig.~2(f)].

\begin{figure*}[t]
\centering
\includegraphics[width=\textwidth]{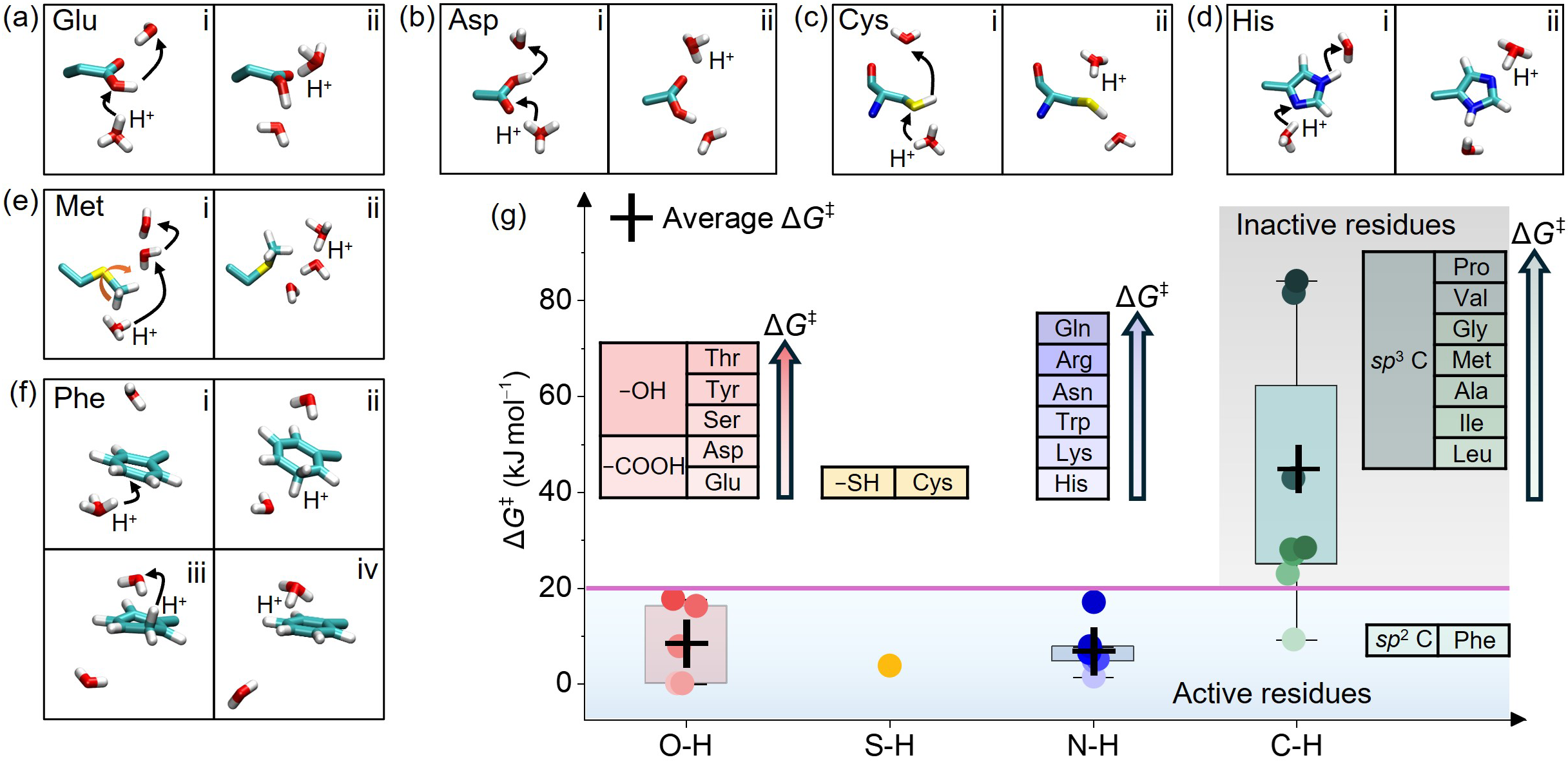}
\caption{
The activity of amino acid residues determined by the \(\mathrm{X}\)--H bond switching energy barrier. 
(a--f) Snapshots of representative proton transfer processes in Glu (a), Asp (b), Cys (c), His (d), Met (e), and Phe (f) residues. Glu, Asp, Cys, His, and Phe mediate proton relay via X–H bond switching, whereas Met supports proton transfer through the adjacent water wire.
(g) Energy barriers \(\Delta G^{\ddagger}\) for \(\mathrm{X}\)--H bond switching in different residues. Residues with \(\Delta G^{\ddagger}\) below \(20~\mathrm{kJ~mol^{-1}}\) are classified as active residues for proton relay, whereas those with \(\Delta G^{\ddagger}\) above \(20~\mathrm{kJ~mol^{-1}}\) are considered as inactive residues.
}
\label{fig:fig2}
\end{figure*}

To quantitatively evaluate the X--H bond switching activity of the 20 amino acid residues, we utilized the extended adaptive biasing force (eABF) method (see SM Sec.~S2~\cite{SM} for details) to calculate the free energy barriers of X--H bond switching (\(\Delta G^{\ddagger}\)) on different residues. As no spontaneous C\textit{sp}\textsuperscript{3}--H bond switching process was observed in inactive C\textit{sp}\textsuperscript{3}--H bonded residues, hypothetical proton relay pathways were constructed to estimate the \(\Delta G^{\ddagger}\) of inactive residues (as detailed in SM Sec.~S2~\cite{SM}). The results show that all active residues with different X--H covalent bonds exhibit low \(\Delta G^{\ddagger}\) ($\sim 0$--$17.66~\mathrm{kJ\,mol^{-1}}$) [Fig.~2(g); see Table~S1~\cite{SM} for complete data]. The average \(\Delta G^{\ddagger}\) of X--H bond switching on active residues follows the order S--H ($3.72~\mathrm{kJ\,mol^{-1}}$) $<$ N--H ($7.53~\mathrm{kJ\,mol^{-1}}$) $<$ O--H ($8.33~\mathrm{kJ\,mol^{-1}}$) $<$ C\textit{sp}\textsuperscript{2}--H ($9.20~\mathrm{kJ\,mol^{-1}}$). Specifically, the \(\Delta G^{\ddagger}\) of O--H bonded residues range from $0$ to $17.66~\mathrm{kJ\,mol^{-1}}$, while those of N--H bonded residues range from $4.23$ to $17.03~\mathrm{kJ\,mol^{-1}}$. Notably, among O--H bonded residues, Glu and Asp containing $-\mathrm{COOH}$ groups exhibit nearly barrierless O--H bond switching processes, consistent with their extensive roles in mediating proton relay within proton channels and enzymes~\cite{Luecke1998,Lampret2020,Muhlbauer2020,Durr2021,Pisliakov2008}. In contrast, the inactive residues exhibit significantly higher \(\Delta G^{\ddagger}\), ranging from $22.89$ to $84.10~\mathrm{kJ\,mol^{-1}}$, which greatly suppresses the probability of C\textit{sp}\textsuperscript{3}--H bond switching mediated proton relay.

\begin{figure*}[t]
\centering
\includegraphics[width=\textwidth]{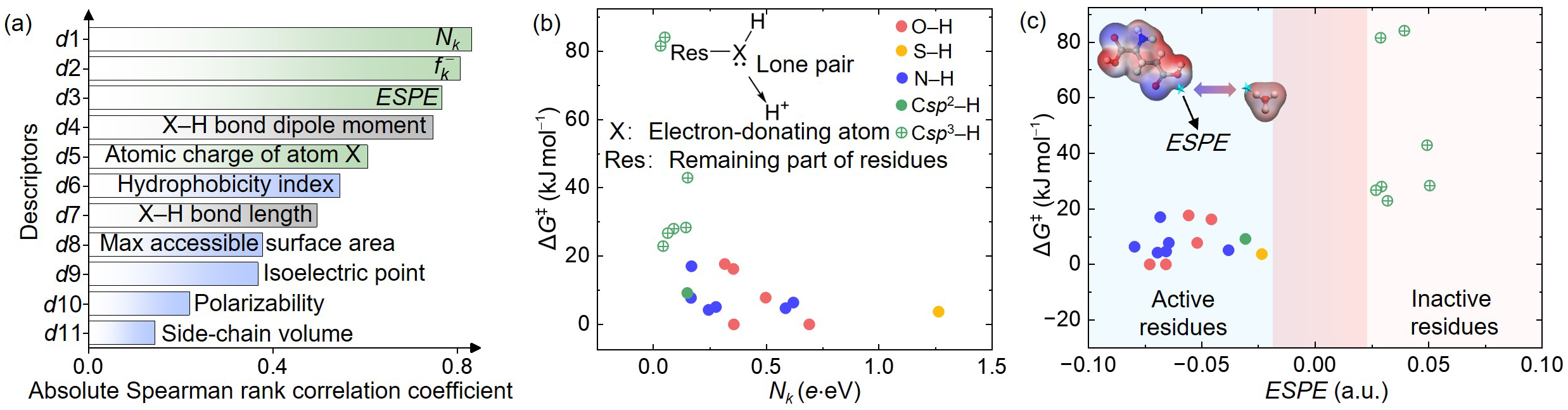}
\caption{
Local nucleophilicity and electrostatic potential extremum of the \(\mathrm{X}\) atom collaboratively regulate the energy barrier of \(\mathrm{X}\)--H bond switching.
(a) Absolute Spearman rank correlation coefficients between 11 descriptors of amino acids and the energy barrier \(\Delta G^{\ddagger}\) of \(\mathrm{X}\)--H bond switching. Descriptors associated with the physicochemical properties of \(\mathrm{X}\) atoms, \(\mathrm{X}\)--H bonds and residues are shown in green, grey and blue, respectively. The local nucleophilicity \((N_k\) and \(f_k^-)\) and \(\mathit{ESPE}\) of the X atoms are the main factors governing \(\Delta G^{\ddagger}\). 
(b) Negative correlation between \(\Delta G^{\ddagger}\) and \(N_k\) of the \(\mathrm{X}\) atoms. 
The inset shows that \(N_k\) characterizes the electron-donating ability of X atom toward the incoming proton, facilitating the formation of a transient X--H bond.
(c) Correlation between \(\Delta G^{\ddagger}\) and \(\mathit{ESPE}\) of the X atoms. Active residues are distributed in the negative \(\mathit{ESPE}\) region, while inactive residues are located in the positive \(\mathit{ESPE}\) region.
The inset shows the electrostatic attraction between amino acids and $\mathrm{H_3O^+}$. The cyan points indicate the \(\mathit{ESPE}\) selected for X atom and $\mathrm{H_3O^+}$.
}
\label{fig:fig3}
\end{figure*}

To identify the key factors regulating \(\Delta G^{\ddagger}\) of \(\mathrm{X}\)--H bond switching, we investigated the absolute Spearman rank correlation between \(\Delta G^{\ddagger}\) and 11 types of physicochemical descriptors that may influence the barrier, including the local nucleophilicity index \((N_k)\), nucleophilic Fukui function \((f_k^-)\), electrostatic potential extremum \((\mathit{ESPE})\), and atomic charge of the proton accepting atom (X atom), and several structural and physicochemical properties of \(\mathrm{X}\)--H bonds and residues [Fig.~3(a) and Figure S4]. 
Results show that \(N_k\), \(f_k^-\), \(\mathit{ESPE}\) and \(\mathrm{X}\)--H bond dipole moment exhibit the largest absolute Spearman rank correlation coefficients with \(\Delta G^{\ddagger}\) [Fig.~3(a)]. 
Specifically, \(N_k\) and \(f_k^-\) both characterize the electron-donating ability of a residue to an incoming proton (see SM Sec.~S3~\cite{SM} for details), thereby quantifying its ability to form a transient \(\mathrm{X}\)--H bond between the X atom and the proton, a step that facilitates the release of the hydrogen atom originally bonded to the X atom. 
We show that there is a negative correlation between \(\Delta G^{\ddagger}\) and \(N_k\), as well as \(f_k^-\) [Figs.~3(b) and Fig.~S4(a)~\cite{SM}]. 
The inactive residues exhibit small \(N_k\) and correspond to high \(\Delta G^{\ddagger}\) values, whereas active residues with large \(N_k\) are mainly located in the low \(\Delta G^{\ddagger}\) region. 
Therefore, residues with larger \(N_k\) possess a stronger ability to stabilize the incoming proton on the X atom and tend to exhibit higher bond switching activity with lower energy barriers. 

Furthermore, active and inactive residues can be clearly demarcated by the \(\mathit{ESPE}\) of their \(\mathrm{X}\) atoms (the $\mathit{ESPE}$ selected for each residue is shown in Fig.~S5~\cite{SM}). 
As shown in Fig.~3(c), all active residues exhibit negative \(\mathit{ESPE}\) \((<-0.02~\mathrm{a.u.})\) and low \(\Delta G^{\ddagger}\), whereas all inactive residues are distributed in the positive \(\mathit{ESPE}\) region \((>0.02~\mathrm{a.u.})\) and exhibit substantially higher \(\Delta G^{\ddagger}\). 
A distinct \(\mathit{ESPE}\) gap between the active and inactive residue regions, ranging from \(-0.02~\mathrm{a.u.}\) to \(0.02~\mathrm{a.u.}\), indicates that \(\mathit{ESPE}\) serves as an effective descriptor for classifying low-barrier active residues and high-barrier inactive residues. 
This result further suggests that a negative local electrostatic environment around the X atom favors the attraction of the positively charged hydronium ion \((\mathrm{H_3O^+})\), thereby facilitating proton approach to the residue and lowering the energy barrier of \(\mathrm{X}\)--H bond switching. Conversely, a positive local electrostatic environment around the \(\mathrm{X}\) atom repels the incoming proton, thereby increasing the energy barrier of \(\mathrm{X}\)--H bond switching. Although the X--H bond dipole moment ranks fourth among the descriptors, it cannot effectively distinguish active and inactive residues (Fig.~S4(b)~\cite{SM}).

Importantly, we further find that the \(\mathrm{X}\)--H bond switching mechanism is not limited to the hHv1 proton channel. The N--H bond switching also occurs at the His292 and Lys282 residues of respiratory complex I~\cite{Sazanov2015}, a key enzyme responsible for biological energy conversion [Figs.~4(a) and 4(b), and Fig.~S6~\cite{SM}]. Further simulations show that the covalent N--H bond switching occurs as well when His292 is mutated to active Asn and Gln residues [Figs.~4(c) and 4(d)].

\begin{figure}[t]
\centering
\includegraphics[width=\columnwidth]{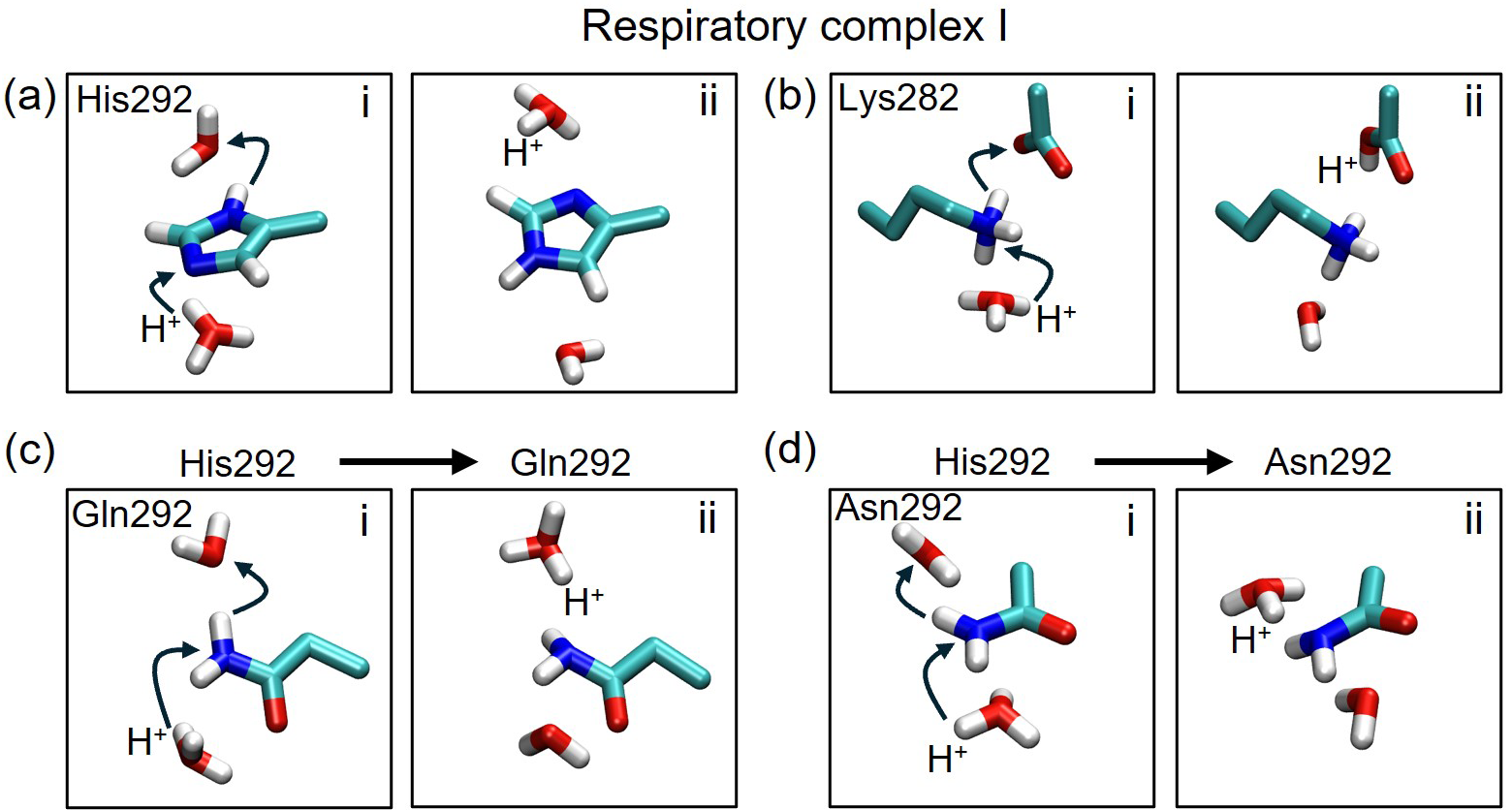}
\caption{
N--H bond switching in respiratory complex I. 
(a,b) Representative snapshots of N--H bond switching processes mediated by His292 (a) and Lys282 (b) residues in respiratory complex I. 
(c,d) The same N--H bond switching occurs after mutation of His292 to Gln292 (c) and Asn292 (d).
}
\label{fig:fig4}
\end{figure}
To further elucidate the function of the distinct activities of amino acids, we analyzed the radial distribution of active and inactive residues within representative protein channels, including hHv1, the C domain of the otopetrin 1 proton channel (Otop1)~\cite{Tu2018} and M2 proton channel of influenza A virus (M2)~\cite{Schnell2008}. In all three channels, active residues tend to be enriched near the central pore region, whereas inactive residues are more frequently distributed toward the peripheral regions of the proteins [Fig.~5(a)]. This result suggests a potential distribution-function relationship that active residues are located along the inner protein region to mediate proton relay through \(\mathrm{X}\)--H bond switching, while inactive residues are located in the outer protein region for structural packing and stability~\cite{Moon2011,Hessa2005}. Specifically, hHv1 generally follows this distribution pattern but exhibits a nonmonotonic radial profile when $r >$ 9 \text{\AA}\ [Fig.~5(b) and Fig. S7(a)~\cite{SM}]. This may arise from its hourglass-like architecture~\cite{Morgan2013}, where active residues along the widened channel entrances are counted at relatively large radial distances~\cite{Jardin2022}. For the C domain of Otop1, a proton channel in sour taste receptor cells, active residues decrease more gradually with increasing \(r\) than in hHv1 and M2, while inactive residues show a similar gradual increase [Fig.~5(c) and Fig.~S7(b)~\cite{SM}]. This broader distribution of residues reflects the complex structure of Otop1 and its multiple proton conduction pathways~\cite{Li2022}. The M2 proton channel in the viral envelope shows a much sharper radial separation between active and inactive residues [Fig.~5(d) and Fig.~S7(c)~\cite{SM}]. This trend is consistent with the simple architecture of M2, in which proton transfer is largely confined to a narrow central pathway ~\cite{Hu2010,Liang2016,Watkins2022}. Overall, these results point to a possible spatial organization principle of amino acids in proton channels: active residues are preferentially positioned near the central regions to mediate the proton relay via X--H bond switching, whereas inactive residues are enriched in peripheral regions to maintain the structural framework of the channel. Nevertheless, the detailed radial distribution of amino acids varies among different proteins, contributing to specific structural features and functions.

\begin{figure}[t]
\centering
\includegraphics[width=\columnwidth]{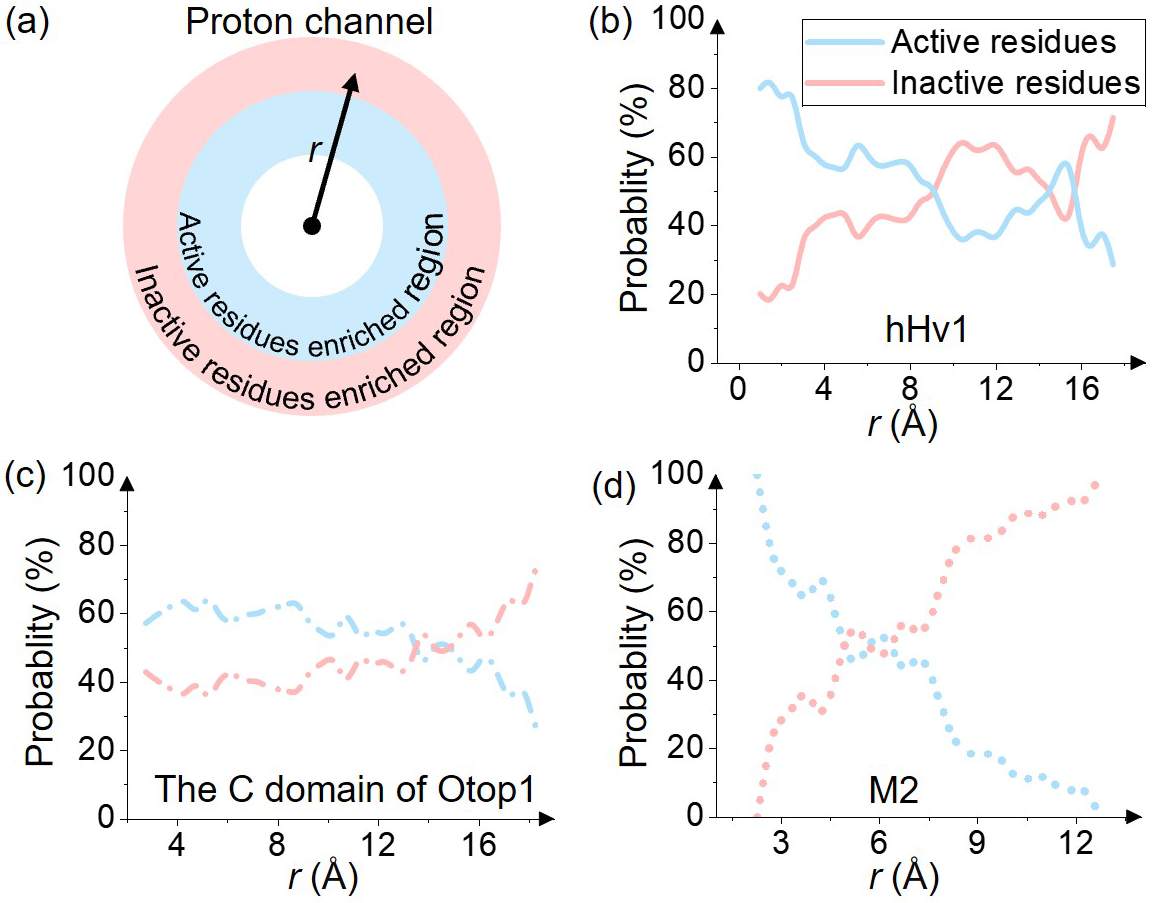}
\caption{
Radial separation of active and inactive residues. (a) A schematic diagram shows that active residues are mainly located along the inner proton conduction regions, whereas inactive residues are more frequently distributed in the outer protein regions. 
(b--d) Radial distributions of active and inactive residues in hHv1 (b), the C domain of Otop1 (c), and M2 (d) proton channels.
}
\label{fig:fig5}
\end{figure}

In summary, we reclassify the twenty canonical amino acids according to their X--H bond switching activity in proton relay, beyond conventional classifications based on charge, acidity/basicity, polarity, or hydrophobicity. The 13 active residues, characterized by terminal O/S/N/C$sp^{2}$--H bonds, mediate proton relay through low barrier X--H bond switching, whereas the remaining 7 residues, characterized by terminal C$sp^{3}$--H bonds, are inactive for bond switching and mainly support proton transfer through adjacent water wires. This activity is governed by the electrostatic potential extremum at the proton accepting atom: active residues exhibit a negative electrostatic potential extremum, while inactive residues exhibit a positive one. We further reveal a distribution-function relationship of amino acids in proton channels, with active residues enriched within the pore to mediate proton relay and inactive residues predominantly distributed in peripheral regions for structural stability. These findings provide new physical insights into how biological systems exploit the twenty canonical amino acids for efficient proton transfer.


\begin{thebibliography}{99}

\bibitem{Crick1961}
F. H. C. Crick, L. Barnett, S. Brenner, and R. J. Watts-Tobin,
General nature of the genetic code for proteins,
Nature \textbf{192}, 1227 (1961).


\bibitem{Machin2023}
J. M. Machin, A. C. Kalli, N. A. Ranson, and S. E. Radford,
Protein--lipid charge interactions control the folding of outer membrane proteins into asymmetric membranes,
Nat. Chem. \textbf{15}, 1754 (2023).

\bibitem{Kamtekar1993}
S. Kamtekar, J. M. Schiffer, H. Xiong, J. M. Babik, and M. H. Hecht,
Protein design by binary patterning of polar and nonpolar amino acids,
Science \textbf{262}, 1680 (1993).

\bibitem{Micheletti1998}
C. Micheletti, F. Seno, A. Maritan, and J. R. Banavar,
Protein design in a lattice model of hydrophobic and polar amino acids,
Phys. Rev. Lett. \textbf{80}, 2237 (1998).

\bibitem{Rose1985}
G. D. Rose, A. R. Geselowitz, G. J. Lesser, R. H. Lee,
and M. H. Zehfus,
Hydrophobicity of amino acid residues in globular proteins,
Science \textbf{229}, 834 (1985).

\bibitem{vanDijk2016}
E. van Dijk, P. Varilly, T. P. J. Knowles, D. Frenkel, and S. Abeln,
Consistent treatment of hydrophobicity in protein lattice models accounts for cold denaturation,
Phys. Rev. Lett. \textbf{116}, 078101 (2016).

\bibitem{Bianco2017}
V. Bianco, G. Franzese, C. Dellago, and I. Coluzza,
Role of water in the selection of stable proteins at ambient and extreme thermodynamic conditions,
Phys. Rev. X \textbf{7}, 021047 (2017).

\bibitem{Dudko2011}
O. K. Dudko, T. G. W. Graham, and R. B. Best,
Locating the barrier for folding of single molecules under an external force,
Phys. Rev. Lett. \textbf{107}, 208301 (2011).

\bibitem{Dill2012}
K. A. Dill and J. L. MacCallum,
The protein-folding problem, 50 years on,
Science \textbf{338}, 1042 (2012).

\bibitem{Yuan2024}
J. Yuan and H. Tanaka,
Impact of hydrodynamic interactions on the kinetic pathway of protein folding,
Phys. Rev. Lett. \textbf{132}, 138402 (2024).

\bibitem{Valet2019}
M. Valet, L.-L. Pontani, R. Voituriez, E. Wandersman, and A. M. Prevost,
Diffusion through nanopores in connected lipid bilayer networks,
Phys. Rev. Lett. \textbf{123}, 088101 (2019).

\bibitem{Gibby2021}
W. A. T. Gibby, M. L. Barabash, C. Guardiani, D. G. Luchinsky,
and P. V. E. McClintock,
Physics of selective conduction and point mutation in biological ion channels,
Phys. Rev. Lett. \textbf{126}, 218102 (2021).

\bibitem{Doyle1998}
D. A. Doyle, J. Morais Cabral, R. A. Pfuetzner, A. Kuo,
J. M. Gulbis, S. L. Cohen, B. T. Chait, and R. MacKinnon,
The structure of the potassium channel: molecular basis of K$^+$ conduction and selectivity,
Science \textbf{280}, 69 (1998).

\bibitem{Kolomeisky2007}
A. B. Kolomeisky,
Channel-facilitated molecular transport across membranes: attraction, repulsion, and asymmetry,
Phys. Rev. Lett. \textbf{98}, 048105 (2007).



\bibitem{Toyabe2010}
S. Toyabe, T. Okamoto, T. Watanabe-Nakayama, H. Taketani,
S. Kudo, and E. Muneyuki,
Nonequilibrium energetics of a single F$_1$-ATPase molecule,
Phys. Rev. Lett. \textbf{104}, 198103 (2010).

\bibitem{Kampjut2020}
D. Kampjut and L. A. Sazanov,
The coupling mechanism of mammalian respiratory complex I,
Science \textbf{370}, eabc4209 (2020).

\bibitem{Morgan2009}
D. Morgan, M. Capasso, B. Musset, V. V. Cherny, E. R{\'i}os,
M. J. S. Dyer, and T. E. DeCoursey,
Voltage-gated proton channels maintain pH in human neutrophils during phagocytosis,
Proc. Natl. Acad. Sci. U.S.A. \textbf{106}, 18022 (2009).

\bibitem{Musset2011}
B. Musset, S. M. Smith, S. Rajan, D. Morgan, V. V. Cherny,
and T. E. DeCoursey,
Aspartate 112 is the selectivity filter of the human voltage-gated proton channel,
Nature \textbf{480}, 273 (2011).

\bibitem{Kim2009}
Y. C. Kim, L. A. Furchtgott, and G. Hummer,
Biological proton pumping in an oscillating electric field,
Phys. Rev. Lett. \textbf{103}, 268102 (2009).

\bibitem{Wu2014}
L. J. Wu,
Voltage-gated proton channel HV1 in microglia,
Neuroscientist \textbf{20}, 599 (2014).

\bibitem{Marochkin2012}
I. I. Marochkin and O. V. Dorofeeva,
Amide bond dissociation enthalpies: effect of substitution on N--C bond strength,
Comput. Theor. Chem. \textbf{991}, 182 (2012).

\bibitem{Roux2010}
M. V. Roux, C. Foces-Foces, R. Notario, M. A. V. Ribeiro da Silva,
M. D. M. C. Ribeiro da Silva, A. F. L. O. M. Santos, and E. Juaristi,
Experimental and computational thermochemical study of sulfur-containing amino acids:
L-cysteine, L-cystine, and L-cysteine-derived radicals. S--S, S--H, and C--S bond dissociation enthalpies,
J. Phys. Chem. B \textbf{114}, 10530 (2010).

\bibitem{Zhou2018}
H. X. Zhou and X. Pang,
Electrostatic interactions in protein structure, folding, binding, and condensation,
Chem. Rev. \textbf{118}, 1691 (2018).

\bibitem{Perrin1997}
C. L. Perrin and J. B. Nielson,
Strong hydrogen bonds in chemistry and biology,
Annu. Rev. Phys. Chem. \textbf{48}, 511 (1997).

\bibitem{Roth1996}
C. M. Roth, B. L. Neal, and A. M. Lenhoff,
Van der Waals interactions involving proteins,
Biophys. J. \textbf{70}, 977 (1996).

\bibitem{Blanksby2003}
S. J. Blanksby and G. B. Ellison,
Bond dissociation energies of organic molecules,
Acc. Chem. Res. \textbf{36}, 255 (2003).

\bibitem{Treyde2022}
W. Treyde, K. Riedmiller, and F. Gr{\"a}ter,
Bond dissociation energies of X--H bonds in proteins,
RSC Adv. \textbf{12}, 34557 (2022).

\bibitem{Hollingsworth2018}
S. A. Hollingsworth and R. O. Dror,
Molecular dynamics simulation for all,
Neuron \textbf{99}, 1129 (2018).

\bibitem{MacKerell1998}
A. D. MacKerell \textit{et al.},
All-atom empirical potential for molecular modeling and dynamics studies of proteins,
J. Phys. Chem. B \textbf{102}, 3586 (1998).

\bibitem{Yukawa2025}
H. Yukawa, H. Kono, H. Ishiwata, R. Igarashi, Y. Takakusagi,
S. Arai, Y. Hirano, T. Suhara, and Y. Baba,
Quantum life science: biological nano quantum sensors, quantum technology-based hyperpolarized MRI/NMR, quantum biology, and quantum biotechnology,
Chem. Soc. Rev. \textbf{54}, 3293 (2025).

\bibitem{Cao2024}
Y. Cao, W. Zhou, C. Shen, H. Qiu, and W. Guo,
Proton Coulomb blockade effect involving covalent oxygen--hydrogen bond switching,
Phys. Rev. Lett. \textbf{132}, 188401 (2024).

\bibitem{Hu2010}
F. Hu, W. Luo, and M. Hong,
Mechanisms of proton conduction and gating in influenza M2 proton channels from solid-state NMR,
Science \textbf{330}, 505 (2010).

\bibitem{Liang2016}
R. Liang, J. M. J. Swanson, J. J. Madsen, M. Hong,
W. F. DeGrado, and G. A. Voth,
Acid activation mechanism of the influenza A M2 proton channel,
Proc. Natl. Acad. Sci. U.S.A. \textbf{113}, E6955 (2016).

\bibitem{Watkins2022}
L. C. Watkins, W. F. DeGrado, and G. A. Voth,
Multiscale simulation of an influenza A M2 channel mutant reveals key features of its markedly different proton transport behavior,
J. Am. Chem. Soc. \textbf{144}, 769 (2022).

\bibitem{SM}
See Supplemental Material at [URL will be inserted by publisher] for
details of the simulation methods of proton transfer, free energy barrier calculations,
descriptor calculations, X--H bond switching snapshots,
electrostatic potential maps, and the descriptor dataset, which includes Refs.~\cite{
Geragotelis2020,Jo2008,Phillips2020,Klauda2010,Jorgensen1983,
Humphrey1996,Melo2018,Stewart1990,Lin2005,Baradaran2013,
Maragliano2006,Sugita1999,Fu2016,Lesage2017,Zamyatnin1972,
Zimmerman1968,Tien2013,Swart2004,Kyte1982,BIOVIA2023,
Gaussian16,Lu2012,Domingo2016CDFT,Chakraborty2021CDFT}.

\bibitem{Geragotelis2020}
A. D. Geragotelis, M. L. Wood, H. G{\"o}ddeke, L. Hong, P. D. Webster, E. K. Wong, J. A. Freites, F. Tombola, and D. J. Tobias,
Voltage-dependent structural models of the human Hv1 proton channel from long-timescale molecular dynamics simulations,
Proc. Natl. Acad. Sci. U.S.A. \textbf{117}, 13490 (2020).

\bibitem{Jo2008}
S. Jo, T. Kim, V. G. Iyer, and W. Im,
CHARMM-GUI: a web-based graphical user interface for CHARMM,
J. Comput. Chem. \textbf{29}, 1859 (2008).

\bibitem{Phillips2020}
J. C. Phillips \textit{et al.},
Scalable molecular dynamics on CPU and GPU architectures with NAMD,
J. Chem. Phys. \textbf{153}, 044130 (2020).

\bibitem{Klauda2010}
J. B. Klauda, R. M. Venable, J. A. Freites, J. W. O'Connor, D. J. Tobias,
C. Mondragon-Ramirez, I. Vorobyov, A. D. MacKerell, Jr., and R. W. Pastor,
Update of the CHARMM all-atom additive force field for lipids: validation on six lipid types,
J. Phys. Chem. B \textbf{114}, 7830 (2010).

\bibitem{Jorgensen1983}
W. L. Jorgensen, J. Chandrasekhar, J. D. Madura, R. W. Impey, and M. L. Klein,
Comparison of simple potential functions for simulating liquid water,
J. Chem. Phys. \textbf{79}, 926 (1983).

\bibitem{Humphrey1996}
W. Humphrey, A. Dalke, and K. Schulten,
VMD: visual molecular dynamics,
J. Mol. Graph. \textbf{14}, 33 (1996).

\bibitem{Melo2018}
M. C. R. Melo, R. C. Bernardi, T. Rudack, M. Scheurer, C. Riplinger,
J. C. Phillips, J. D. C. Maia, G. B. Rocha, J. V. Ribeiro, J. E. Stone,
F. Neese, K. Schulten, and Z. Luthey-Schulten,
NAMD goes quantum: an integrative suite for hybrid simulations,
Nat. Methods \textbf{15}, 351 (2018).

\bibitem{Stewart1990}
J. J. P. Stewart,
MOPAC: a semiempirical molecular orbital program,
J. Comput. Aided Mol. Des. \textbf{4}, 1 (1990).

\bibitem{Lin2005}
H. Lin and D. G. Truhlar,
Redistributed charge and dipole schemes for combined quantum mechanical and molecular mechanical calculations,
J. Phys. Chem. A \textbf{109}, 3991 (2005).

\bibitem{Baradaran2013}
R. Baradaran, J. M. Berrisford, G. S. Minhas, and L. A. Sazanov,
Crystal structure of the entire respiratory complex I,
Nature \textbf{494}, 443 (2013).

\bibitem{Maragliano2006}
L. Maragliano, A. Fischer, E. Vanden-Eijnden, and G. Ciccotti,
String method in collective variables: minimum free energy paths and isocommittor surfaces,
J. Chem. Phys. \textbf{125}, 024106 (2006).

\bibitem{Sugita1999}
Y. Sugita and Y. Okamoto,
Replica-exchange molecular dynamics method for protein folding,
Chem. Phys. Lett. \textbf{314}, 141 (1999).

\bibitem{Fu2016}
H. Fu, X. Shao, C. Chipot, and W. Cai,
Extended adaptive biasing force algorithm. An on-the-fly implementation for accurate free-energy calculations,
J. Chem. Theory Comput. \textbf{12}, 3506 (2016).

\bibitem{Lesage2017}
A. Lesage, T. Leli{\`e}vre, G. Stoltz, and J. H{\'e}nin,
Smoothed biasing forces yield unbiased free energies with the extended-system adaptive biasing force method,
J. Phys. Chem. B \textbf{121}, 3676 (2017).

\bibitem{Zamyatnin1972}
A. A. Zamyatnin,
Protein volume in solution,
Prog. Biophys. Mol. Biol. \textbf{24}, 107 (1972).

\bibitem{Zimmerman1968}
J. M. Zimmerman, N. Eliezer, and R. Simha,
The characterization of amino acid sequences in proteins by statistical methods,
J. Theor. Biol. \textbf{21}, 170 (1968).

\bibitem{Tien2013}
M. Z. Tien, A. G. Meyer, D. K. Sydykova, S. J. Spielman, and C. O. Wilke,
Maximum allowed solvent accessibilities of residues in proteins,
PLoS ONE \textbf{8}, e80635 (2013).

\bibitem{Swart2004}
M. Swart, J. G. Snijders, and P. T. van Duijnen,
Polarizabilities of amino acid residues,
J. Comput. Methods Sci. Eng. \textbf{4}, 419 (2004).

\bibitem{Kyte1982}
J. Kyte and R. F. Doolittle,
A simple method for displaying the hydropathic character of a protein,
J. Mol. Biol. \textbf{157}, 105 (1982).

\bibitem{BIOVIA2023}
Dassault Syst{\`e}mes BIOVIA,
BIOVIA Materials Studio 2023,
Dassault Syst{\`e}mes (2023).

\bibitem{Gaussian16}
M. J. Frisch \textit{et al.},
Gaussian 16, Revision C.01,
Gaussian, Inc., Wallingford CT (2016).

\bibitem{Lu2012}
T. Lu and F. Chen,
Multiwfn: a multifunctional wavefunction analyzer,
J. Comput. Chem. \textbf{33}, 580 (2012).

\bibitem{Domingo2016CDFT}
L. R. Domingo, M. R\'{\i}os-Guti\'errez, and P. P\'erez,
Applications of the conceptual density functional theory indices to organic chemistry reactivity,
Molecules \textbf{21}, 748 (2016).

\bibitem{Chakraborty2021CDFT}
D. Chakraborty and P. K. Chattaraj,
Conceptual density functional theory based electronic structure principles,
Chem. Sci. \textbf{12}, 6264 (2021).


\bibitem{Luecke1998}
H. Luecke, H.-T. Richter, and J. K. Lanyi,
Proton transfer pathways in bacteriorhodopsin at 2.3 angstrom resolution,
Science \textbf{280}, 1934 (1998).

\bibitem{Lampret2020}
O. Lampret, J. Duan, E. Hofmann, M. Winkler, F. A. Armstrong,
and T. Happe,
The roles of long-range proton-coupled electron transfer in the directionality and efficiency of [FeFe]-hydrogenases,
Proc. Natl. Acad. Sci. U.S.A. \textbf{117}, 20520 (2020).

\bibitem{Muhlbauer2020}
M. E. M{\"u}hlbauer, P. Saura, F. Nuber, A. Di Luca, T. Friedrich,
and V. R. I. Kaila,
Water-gated proton transfer dynamics in respiratory complex I,
J. Am. Chem. Soc. \textbf{142}, 13718 (2020).

\bibitem{Durr2021}
S. L. D{\"u}rr, O. Bohuszewicz, D. Berta, R. Suardiaz,
P. G. Jambrina, C. Peter, Y. Shao, and E. Rosta,
The role of conserved residues in the DEDDh motif: the proton-transfer mechanism of HIV-1 RNase H,
ACS Catal. \textbf{11}, 7915 (2021).

\bibitem{Pisliakov2008}
A. V. Pisliakov, P. K. Sharma, Z. T. Chu, M. Haranczyk,
and A. Warshel,
Electrostatic basis for the unidirectionality of the primary proton transfer in cytochrome c oxidase,
Proc. Natl. Acad. Sci. U.S.A. \textbf{105}, 7726 (2008).

\bibitem{Sazanov2015}
L. A. Sazanov,
A giant molecular proton pump: structure and mechanism of respiratory complex I,
Nat. Rev. Mol. Cell Biol. \textbf{16}, 375 (2015).

\bibitem{Tu2018}
Y.-H. Tu, A. J. Cooper, B. Teng, R. B. Chang, D. J. Artiga,
H. N. Turner, E. M. Mulhall, W. Ye, A. D. Smith, and E. R. Liman,
An evolutionarily conserved gene family encodes proton-selective ion channels,
Science \textbf{359}, 1047 (2018).

\bibitem{Schnell2008}
J. R. Schnell and J. J. Chou,
Structure and mechanism of the M2 proton channel of influenza A virus,
Nature \textbf{451}, 591 (2008).

\bibitem{Moon2011}
C. P. Moon and K. G. Fleming,
Side-chain hydrophobicity scale derived from transmembrane protein folding into lipid bilayers,
Proc. Natl. Acad. Sci. U.S.A. \textbf{108}, 10174 (2011).

\bibitem{Hessa2005}
T. Hessa, H. Kim, K. Bihlmaier, C. Lundin, J. Boekel, H. Andersson,
I. Nilsson, S. H. White, and G. von Heijne,
Recognition of transmembrane helices by the endoplasmic reticulum translocon,
Nature \textbf{433}, 377 (2005).

\bibitem{Morgan2013}
D. Morgan, B. Musset, K. Kulleperuma, S. M. E. Smith, S. Rajan,
V. V. Cherny, R. Pom\`es, and T. E. DeCoursey,
Peregrination of the selectivity filter delineates the pore of the human voltage-gated proton channel hHV1,
J. Gen. Physiol. \textbf{142}, 625 (2013).

\bibitem{Jardin2022}
C. Jardin, N. Ohlwein, A. Franzen, G. Chaves, and B. Musset,
The pH-dependent gating of the human voltage-gated proton channel from computational simulations,
Phys. Chem. Chem. Phys. \textbf{24}, 9964 (2022).

\bibitem{Li2022}
B. Li, Y. Wang, A. Castro, C. Ng, Z. Wang, H. Chaudhry,
Z. Agbaje, G. A. Ulloa, and Y. Yu,
The roles of two extracellular loops in proton sensing and permeation in human Otop1 proton channel,
Commun. Biol. \textbf{5}, 1110 (2022).

\end{thebibliography}
\end{document}